\pdfoutput=1
\documentclass{du-journals}
\journal{Iranian Journal of Astronomy and Astrophysics}
\title{Structure of proper motions in a sunspot penumbra\footnote{Published in IJAA, 2015, Vol. 2, Issue 1, Pages 39-57, DOI: 10.22128/IJAA.2015.16 (https://ijaa.du.ac.ir/).}}
\author[1]{H. Hamedivafa}
\address[1]{Physics Department, Faculty of Science, Imam Khomeini International University,
Qazvin 34149-16818, Islamic Rep. of Iran; email:
vafa@sci.ikiu.ac.ir}
\begin{document}
\begin{abstract}
We study the structure and evolution of the horizontal proper
motions in a regular sunspot penumbra, very close to the solar disc
center, in active region NOAA 11092 using a 48~min time sequences of
blue continuum images recorded by \textit{Hinode}/SOT in 2010 August
3. During the day of the observation, the active region had a low
activity level. We apply local correlation tracking (LCT) to the
time series of the images to obtain the field of horizontal proper
motions (flow field). The penumbra shows a slow (fast) flow field
with an average speed of 0.2 (0.4)~km/s starting at its middle
towards the umbra (outer penumbral boundary) as an inward (outward)
motion in accordance with previous findings. This behavior defines a
continuous \textit{divergence line} at the middle of the penumbra
($r\simeq2R_{spot}/3$). A distorted \textit{ringlike} feature with
very slow flows ($\approx50$~m/s; \textit{zero-flow ring}: ZFR)
co-spatial with the divergence line is clearly seen on the speed map
of proper motions in logarithmic scale. Deep intrusion of
coordinated penumbral filaments into the umbra can cause the
ZFR a) to be significantly displaced towards the umbra (in most cases)
or b) to be discontinuous, showing considerable speeds there
($\approx150$~m/s). Where the ZFR shows discontinuity, the
divergence line does not move toward the umbra. Also, because of the
different evolutionary flows of adjacent penumbral filaments, the
ZFR and the divergence line show a stable backward/forward
displacement along itself during the 48~min observation. The radial
variations of the azimuthally averaged brightness show a local
bright ring with a weak contrast of $1\%$ close to the ZFR. At the
outer penumbra, we find that the converging filamentary flow occurs
in a dark radial channel and the filamentary diverging flows are
formed by the evolution of thin bright fibrils. Also, the large
speeds at the penumbra boundary are produced by the displacement
and/or the fragmentation of the bright fibrils in developing
filamentary flows. At the near vicinity of the penumbra, in
surrounding granulation, some divergence centers are strongly pushed
away as a whole with an average speed of about 0.6~km/s by these
developing filamentary flows.
\end{abstract}


\section{Introduction}\label{intro}
Magnetic fields on the Sun interact in various ways with plasma and
radiation producing a large variety of phenomena. In the convection
zone, large and small scale magnetic fields are generated and
partially transferred into the outer layers of the Sun. A sunspot as
seen in the photosphere is the most prominent example of magnetic
phenomena whose magnetic field extends into both the interior and
the outer layers of the Sun. Therefore, sunspots are central to our
understanding of solar magnetism.

At an insufficient spatial resolution (worse than 1~arcsec), the
dark central/inner part of a sunspot, so-called umbra, is surrounded
by a fairly uniform region, so-called penumbra, which is on average
brighter than the umbra, but less bright than the surrounding
granulation. However, at a spatial resolution of a few tenth of
arcsec, a sunspot penumbra exhibits bright and dark filamentary
structures and penumbral grains often appear at the head (the umbral
side) of the bright filaments. Furthermore, it has been demonstrated
that the formation of the penumbra is coupled to the Evershed flow -
a nearly horizontal and radial outflow of plasma from inner penumbra
towards the outer boundary \cite{ever09} - which is correlated to
the filamentary structures of the magnetic field (e.g.,
\cite{title93,leka98}; for a review see \cite{bor_ich11}).

Borrero \& Ichimoto \cite{bor_ich11} (also see references therein)
have presented the spatial correlation between penumbral filaments
and the Evershed flow and obtained that the Evershed flow correlates
with more horizontal magnetic fields throughout the entire penumbra,
while it correlates with bright filaments in the inner penumbra but
with dark filaments in the outer penumbra.

At high angular resolution, penumbral filaments exhibit a central
dark lane (core) and two lateral brightenings \cite{scharmer02}. The
common occurrence of dark-cored filaments and the fact that their
various parts show a coherent behavior have raised expectations that
they could be the fundamental constituents of the penumbra. The
high-resolution spectroscopic measurements (e.g.,
\cite{bell05,bell07,langh07}) have demonstrated that the Evershed
flow appears preferentially in dark cores, although its velocity
varies from one dark-cored filament to another.

The \textit{steady siphon-flow} (e.g.,
\cite{meyer68,montes93,montes97,bor05}) and the \textit{rising
flux-tube} \cite{schlich98a,schlich98b,schlich02} can give two
possible mechanisms of the Evershed flow in the context of an
\textit{uncombed flux-tube} penumbra \cite{solanki93} or
\textit{interlocking comb} structure \cite{thomas92}. In the
\textit{field-free gap model} of a penumbra \cite{scsp06,spsc06},
overturning convection takes place inside field-free gaps within the
penumbra. Then, a radial outflow resembling the Evershed flow may be
driven.

Deng et al. \cite{deng07} point that a phenomenon related to a
change in the local flow field in and around a sunspot, magnetic
flux can affect the formation and decay of the sunspot penumbra.
Therefore, a careful study of flow field in and around a sunspot
penumbra in different evolutionary phases is necessary to understand
the processes that might contribute to sunspot evolution.

In the present work, we do not deal with spectroscopic analysis.
But, we are interested in local horizontal flows in a sunspot
penumbra that can be derived with local correlation tracking (LCT,
\cite{nov_sim88}) applied to a time series of intensity images
(Sect.~\ref{lct}).

Generally, the measurement of the displacement of some tracers,
using broad-band imagery provides two components of the local
horizontal flow. The differential rotation of the Sun was discovered
by tracking sunspots across the solar disc. Also, local horizontal
flows (proper motions) in the solar photosphere can be inferred from
the tracking of solar granulation. In this latter measurement, the
horizontal flows are inferred from the displacement of a bright
feature (\textit{feature tracking}; e.g., \cite{tk07}, \cite{ham11}
and references therein).

However, the LCT algorithm has been used by several authors to
derive the horizontal flow fields in different types of solar
applications, e.g., Brandt et al. \cite{brandt88} found vortex flows
in the solar granulation, November \cite{nov89} studied the proper
motion of solar granulation to search for mesogranulation. Also,
Wang et al. \cite{wang95} studied the vorticity and divergence in
the solar photosphere and Roudier et al. \cite{roud98} measured the
lifetime of solar mesogranules. Bonet et al. \cite{bonet05}, and
Sobotka \& Roudier \cite{sob_roud07} studied the dynamics of moat
flow around sunspots. Molowny-Horas \cite{molo94}, M\'arquez et al.
\cite{marq06} and Denker et al. \cite{denker08} measured horizontal
proper motions in a sunspot penumbra.

We follow the definitions of Denker et al. \cite{denker08} and use
the terms \textit{flow}, \textit{proper motion} or \textit{speed}
when we refer to LCT results and leave the term \textit{velocity}
for spectroscopic/Doppler measurements of plasma motions, although
this latter one is not the case of the present work.

LCT technique was used by Tan et al. \cite{tan09} to study
horizontal proper motions in the penumbra of a rapidly rotating
$\delta$-sunspot using \textit{Hinode} G-band images with 2~min
cadence. They used the Stokes~\textit{V} observations (Fe~\texttt{I}
at 630.2~nm: from \textit{Hinode} Narrowband Filter Imager) to
distinguish the proper motions of positive and negative magnetic
elements in the shear flow region along the magnetic neutral line in
the penumbra. They discussed that the variations of the penumbral
intensity and penumbral flow (mean speeds temporally and spatially
varied from 0.6 to 1.1~km/s within the penumbral decay areas) were
associated with the CME/flare eruptions. Especially, they found that
the magnitude of the shear flow apparently dropped down from 0.6 to
0.3~km/s in response to the magnetic energy release. They applied a
time window of 20~min to average motions (see Sect.~\ref{lct}).

The motion of penumbral grains can affect directly on the LCT flow
maps. It is useful to distinguish between penumbral grains located
in the inner penumbra with lifetimes up to 3 h and penumbral grains
located in the outer penumbra with lifetimes generally less than 1 h
\cite{mull73,sob99a,zha_ich13}. Moreover, the inner penumbral grains
generally move radially inwards at an average speed of 0.3-0.7~km/s
while the outer ones migrate both outwards and inwards at a higher
speed of 0.4-0.9~km/s on average (e.g., \cite{sob99a,zha_ich13,ss01}
and their references).

\section{Local Correlation Tracking Algorithm}\label{lct}

The LCT algorithm which is the method used in the present work
assumes that an ``intensity pattern" is moved by the solar flow
field \cite{nov_sim88,nov89}. Then, horizontal flow field can be
measured by finding the local correlation (called cross-correlation)
of the contrast changes of intensity pattern in two successive
sub-images (correlation windows) which are shifted with respect to
each other. The shift which maximizes the defined cross-correlation
is taken as the ``true" displacement of the intensity structure.

Therefore, using the horizontal flow field (LCT flow maps), we can
find and follow the dynamical structures of the flow. Also,
downflows and upflows can be searched by finding convergence and
divergence areas in LCT flow maps, and we can search for vortices as
well by choosing suitable LCT free parameters.

Nevertheless, it should be noted that the displacements of an
intensity pattern or a bright feature are not necessarily indicative
of real plasma flows and may be due to heat diffusion, traveling
waves or other effects leading to a misinterpretation. On the other
hand, LCT does not distinguish between displacements of the bright
and dark features. For example, Wang \& Zirin \cite{wz92} have
reported the motion of dark fibrils towards the umbra in the inner
part of a penumbra.

A detailed description of this algorithm (LCT) has been reviewed by,
e.g., \v{S}vanda \cite{svanda07} and Verma \& Denker \cite{verma11}.
The former author has processed MDI Dopplergrams measured in the
Dynamics campaigns using the LCT algorithm.

Roudier et al. \cite{roud99} have discussed and compared the two
techniques of LCT and ``feature tracking technique" for
determination of horizontal flow field in the solar granulation.
They suggested that the tracking of ``coherent structures" in binary
images gives more reliable results than using LCT of intensity
pattern. This algorithm was developed by Rieutord et al.
\cite{rieutord07}. This algorithm needs the possibility of selecting
specific structures (e.g. granules, magnetic bright points,
penumbral grains, and umbral dots) in a sequence of images. However,
the image processing steps identifying desired features rely on
prior knowledge about them.

On the other hand, in LCT technique the choice of tracking
parameters such as the size of the \textit{correlation window} which
smoothes the flows and is related to the spatial resolution of
observation, the \textit{time cadence} (lag) which defines a
detectable minimum/maximum speed, and the \textit{time window} over
which a flow map is averaged can significantly affect the result.

However, Verma \& Denker \cite{verma11} have discussed the
optimization and justifying the tracking parameters for LCT in
time-series of G-band images obtained by \textit{Hinode}/SOT with
spatial resolution of 0.11~arcsec/pixel: features with low
displacement rates (speeds) cannot be accurately tracked if the
cadence is not enough long, whereas in much longer cadences,
features will evolve too much so that the algorithm might not
recognize them anymore. Averaging over time scales significantly
longer than the lifetime of individual features can smooth their
evolutionary proper motions resulting in a `global" flow field in a
sequence of images of granulation. They found that the LCT algorithm
produces the best results for G-band images having a cadence of
60-90~s.

Generally, the LCT technique underestimates the real speeds (amount
to $20-30\%$) due to the smoothing of processed data by the
correlation window (e.g., \cite{svanda07,sob99b,rieutord01}) and the
time range which in it LCT technique is applied (see
Sect.~\ref{zfr}).

\section{Data Set}\label{data}

We implemented LCT method to measure horizontal flow field using a
time-series of broad-band images (blue continuum, 450.4~nm) obtained
with the Solar Optical Telescope on board \textit{Hinode} (SOT,
\cite{tsu08,sue08}). Data from space do not suffer the distortion
effects of Earth's turbulent atmosphere so that intensity features
can be followed from image to image.

This data set was acquired from 13:07 to 13:55~UT on August 3, 2010
with a cadence of 60~s. This selected time-series contains 49 images
with the size of $470\times470$ pixels and spatial sampling of
0.11~arcsec/pixel. The spatial resolution of the observation in blue
continuum is around 0.25~arcsec. The observations were centered on
active region NOAA 11092 practically located at disk centre, at
heliocentric coordinates N$20^{\prime\prime}$~W$03^{\prime\prime}$.
The active region has a sunspot with a regular magnetic field
configuration and is classified as a $\beta/\alpha$-region with a
very low activity.

To align the images, we divided the whole series into 7 subseries.
Then, we calculated shifts between current images and a reference
image by computing the linear correlation coefficient and using the
properties of Fourier transforms only on the central part of the
images (mostly umbra)  containing $256\times256$~pixels. These
shifts are then applied in the subseries to align images with
respect to the reference image with sub-pixel accuracy. Exactly
after co-aligning each subseries, its last image was selected as the
reference image for the alignment of the next subseries, except for
the first subseries for which the first image was selected as the
reference image.

\begin{figure} [p!] 
 \centerline{\includegraphics[width=\textwidth]{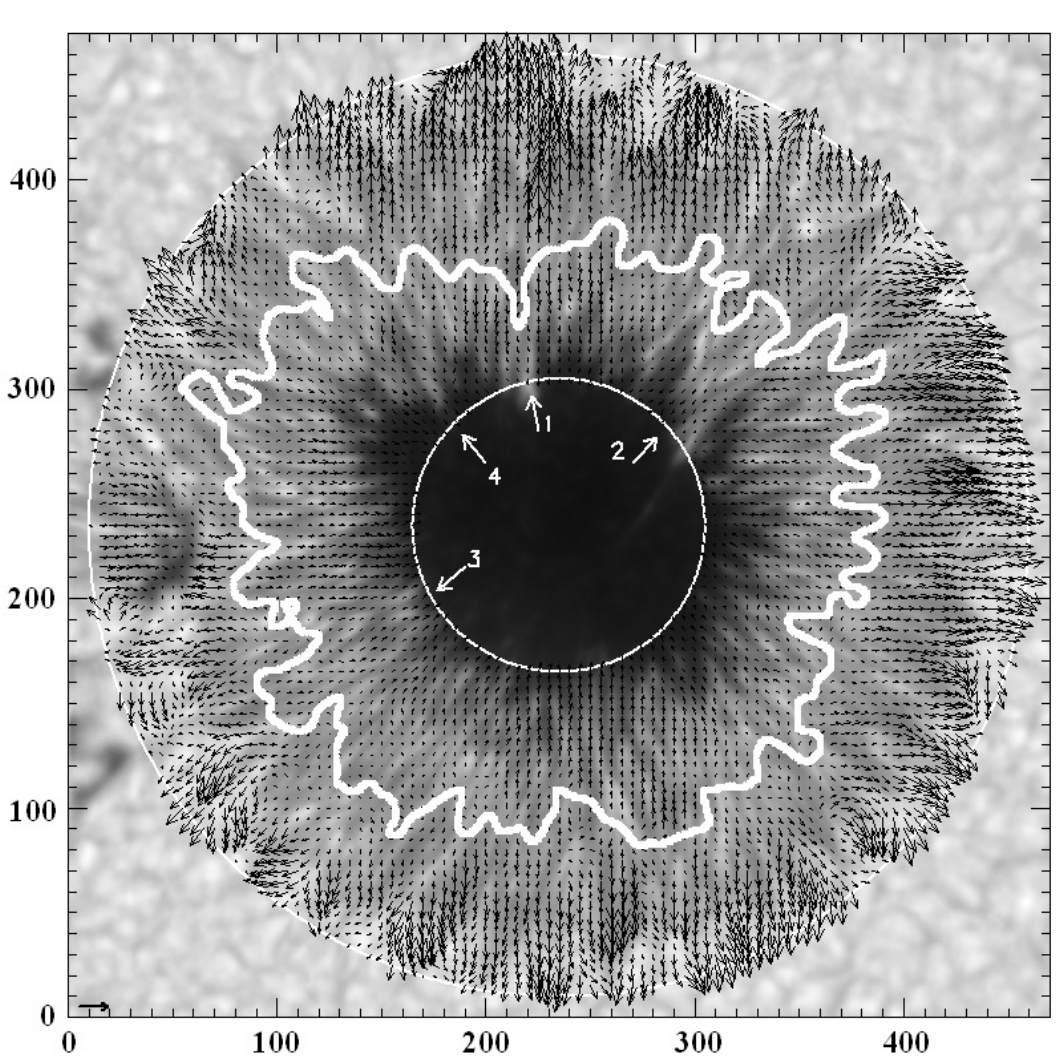}}
 \caption[]{Global vector flow field averaged over the whole period of the time series.
  The FWHM of the Gaussian-weighted correlation window is 1~arcsec.
  The background is the time-averaged intensity map.
  White circles are the defined inner and outer borders of the penumbra.
  White contour encloses the inner penumbra with inward flows.
  Some parts of the umbra and the surrounding granulation are also included in between the two defined borders.
  The black horizontal arrow at the lower left corner is equivalent to 1~km/s.}
 \label{gvecflow}
\end{figure}

\section{Results}\label{res}

As mentioned before, the core of the LCT algorithm is the
cross-correlation function. We used the modified program
\verb"flowmaker.pro" (implemented in IDL \cite{molo_yi94}) in which
we adopted the cross-correlation function as the sum of square of
the local intensity differences in two subsequent images computed
over a Gaussian-weighted correlation window with a FWHM of 1~arcsec
(equivalent to 9~pixels). The position of the maximum of the
cross-correlation values is calculated with sub-pixel accuracy by a
parabola fit to the neighboring pixels. The time lag between
correlated images is 60~s.

\subsection{Flow Maps}\label{flowmaps}

At first, we obtain the global flow field in the sunspot penumbra.
Fig.~\ref{gvecflow} displays the calculated horizontal proper
motions (vector flow field) averaged over the whole period of the
time series, 48~min. The background of the vector flow field is the
averaged intensity map in which stable dark and bright \textit{vena}
(filaments) are seen all over the penumbra. White circles are the
defined inner and outer borders of the penumbra. These borders are
defined so that two annular areas of 2-3~arcsec width of the umbra
and the surrounding granulation are included in all flow maps. Also,
we can see local phenomena in the vector flow field shown in
Fig.~\ref{gvecflow}, e.g. diverging/converging flow filaments in the
outer penumbra and unidirectional inward and outward flows in the
inner and outer penumbra, respectively, and more strongly at the
periphery of the penumbra.

Our analysis confirms previous findings of radial proper motions and
inward and outward flows which depend on the distance to the
penumbral border: a slow flow (up to 0.7~km/s with an average of
0.2~km/s) is starting at the middle of the penumbra towards the
umbra (inward motion) and a fast flow (up to 2.0~km/s with an
average of 0.5~km/s) towards the outer penumbra boundary (outward
motion). This diverging motion defines a \textit{divergence line}
approximately in the middle of the penumbra, in accordance with
\cite{deng07,molo94,denker08}. This divergence line is displayed in
Fig.~\ref{gvecflow} as a white contour overlaid the mean vector flow
field separating the areas with inward motions from those with
outward motions. The distorted area at the left side of the penumbra
was excluded. Also, Sobotka et al. \cite{sob99a} have found a
divergence line in the penumbra when they studied the proper motions
of penumbral grains using feature tracking technique.

\begin{figure} [p!] 
 \centerline{\includegraphics[width=\textwidth,clip=]{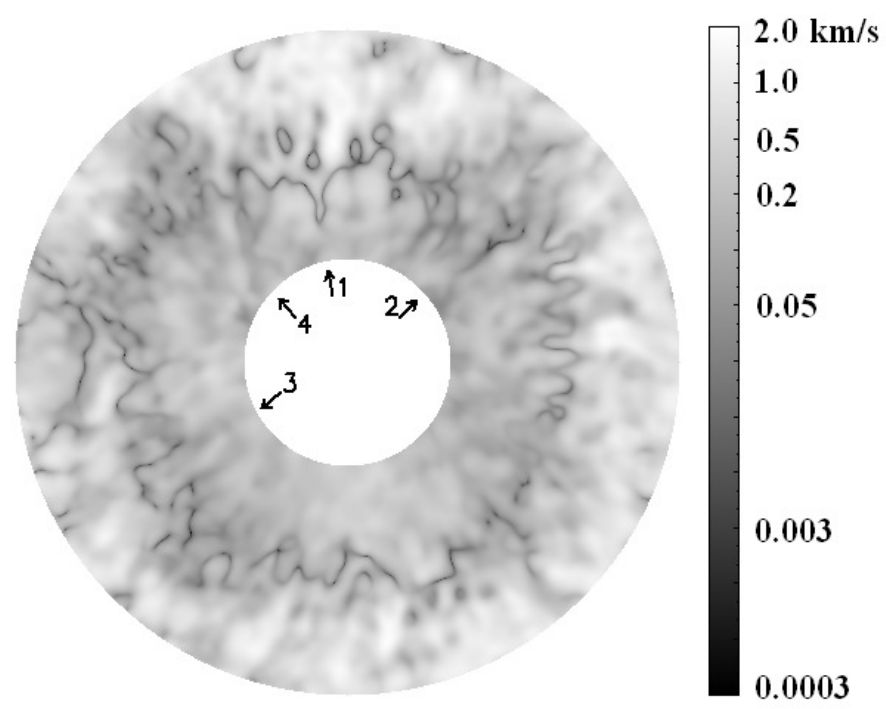}}
 \caption[]{Map of the speed of proper motions (the magnitude of the vector flow field shown in Fig.~\ref{gvecflow}) in logarithmic scale.
}
 \label{gspeed}
\end{figure}

Fig.~\ref{gspeed} shows the speed map displaying the magnitude of
the vector flow field in logarithmic scale. The precision of the
speeds can be estimated to be less than 0.03~km/s, if we compare the
two speed maps obtained from two 48-frames time series which are
shifted only one frame (the first 48~subsequent images and the
sequence of images of~$2-49$). If the speed map is calculated using
a wider correlation window, the obtained proper motions (speeds)
become slower with more smooth pathways (not shown here).

\begin{figure} 
 \centerline{\includegraphics[width=\textwidth,clip=]{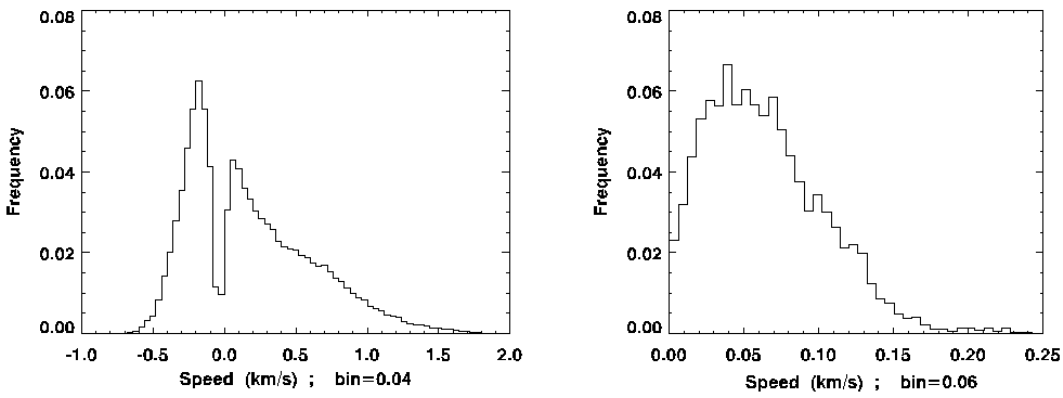}}
 \caption[]{Left panel: histogram of speeds in inner (negative values; inward) and outer (positive vales; outward) penumbra.
  Right panel: histogram of speeds on the divergence line.
 }
 \label{histspeed}
\end{figure}

The histogram of speeds in inner (negative values; inward) and outer
(positive values; outward) penumbra is shown in the left panel in
Fig.~\ref{histspeed}. We can see that the penumbra shows an active
kinematics in the LCT map, because of its low population at zero
speed. This speed distribution is similar to the histogram of inward
and outward speeds obtained by Molowny-Horas \cite{molo94}.

A \textit{ringlike} feature with very slow flows ($\approx50$~m/s in
average; \textit{zero-flow ring}: ZFR) is clearly evident in the
middle of the penumbra on the speed map in logarithmic scale in
which this ring has the best contrast. Since the LCT algorithm makes
a smoothed average of proper motions in a relatively wide
correlation window, at the position of the divergence line, the
proper motions in opposite directions cancel out each other.
Therefore, we find a ZFR on the speed map that must be co-spatial
with the divergence line. The right panel in Fig.~\ref{histspeed}
shows the histogram of speeds on the divergence line. However, the
ZFR is not a smoothed/continuous circle likely due to the different
evolutionary proper motions of penumbral filaments (see
Sect.~\ref{timeevol}). Especially, a wide discontinuity and a
considerable displacement of the ZFR towards the umbra are evident
in regions where coordinated penumbral filaments are deeply
intruding into the umbra. In Figs.~\ref{gvecflow} \& \ref{gspeed},
arrows~1~\&~2 show examples of the displacement of the ZFR and its
discontinuity in the studied penumbra, respectively.

Discontinuities in the ZFR means that there is no ``zero" (very
small) speed there, although we can find a divergence line in those
areas. The averaged speed along the divergence line at the positions
of discontinuities is about 150 m/s which is about three times the
averaged speed in other positions of the divergence line, see the
speed distribution on the divergence line shown in right panel in
Fig.~\ref{histspeed}.

\subsection{Radial Variations}\label{radvar}

\begin{figure} 
 \centerline{\includegraphics[width=\textwidth,clip=]{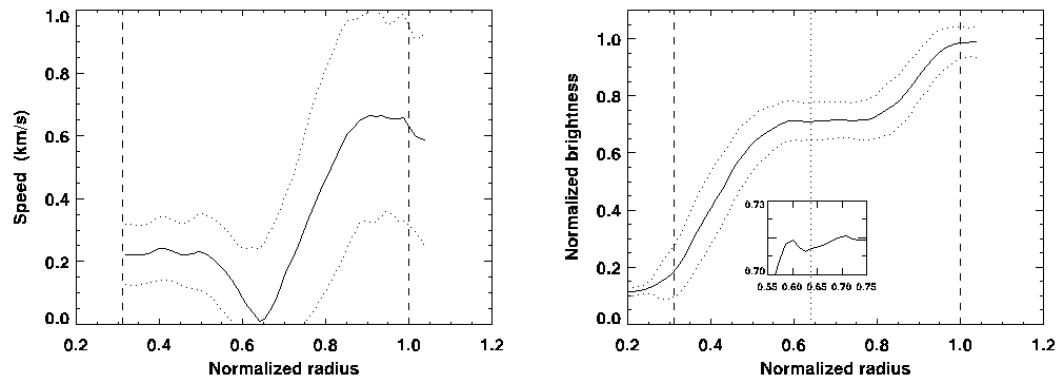}}
 \caption[]{Radial variations of the azimuthally averaged speed (left) and brightness (right).
  The vertical dashed lines mark the penumbral boundaries in both panels.
  The dotted curves in both panels represent the range of standard deviation of the corresponding quantity found at each radial distance.
  The vertical dotted line in the right panel marks the radial position of the minimal speed (radius of ZFR) found from the left panel.
  The inset in the right panel is the magnified profile around $r/R_{spot}=0.6$ showing a weak bright ring in the penumbra.
 }
 \label{IV_vs_r}
\end{figure}

Although the spot is circular, the flow field does not have a
circular symmetry and behaves differently in different azimuthal
directions (non-isotropic). However, we compute the azimuthal
averages of speed and brightness as a function of radius in the
penumbra. Fig.~\ref{IV_vs_r} displays the variation of these two
quantities versus normalized distance from the sunspot centre. The
radial position of the averaged ZFR ($0.65R_{spot}$: its averaged
radius) is in agreement with the averaged radius of the divergence
line obtained by others, e.g. see \cite{denker08} and the references
therein.

The right panel in Fig.~\ref{IV_vs_r} shows a wide plateau region in
the radial profile of the normalized intensity of the studied
penumbra at around 0.7 from 0.6 to $0.8R_{spot}$ which is consistent
with the findings of Westendorp Plaza et al. \cite{plaza01}.

Also, a weak bright ring with a low contrast of 1\% can be seen in
the radial intensity profile at around $r/R_{spot}=0.6$, similar to
as reported in \cite{plaza01}. The inset in the right panel of
Fig.~\ref{IV_vs_r} shows the magnified intensity profile around the
weak bright ring.

We did not find any considerable correlations between intensity and
inward/outward speed.

\subsection{Time Evolution}\label{timeevol}

\begin{figure} [p!] 
 \centerline{\includegraphics[width=\textwidth,clip=]{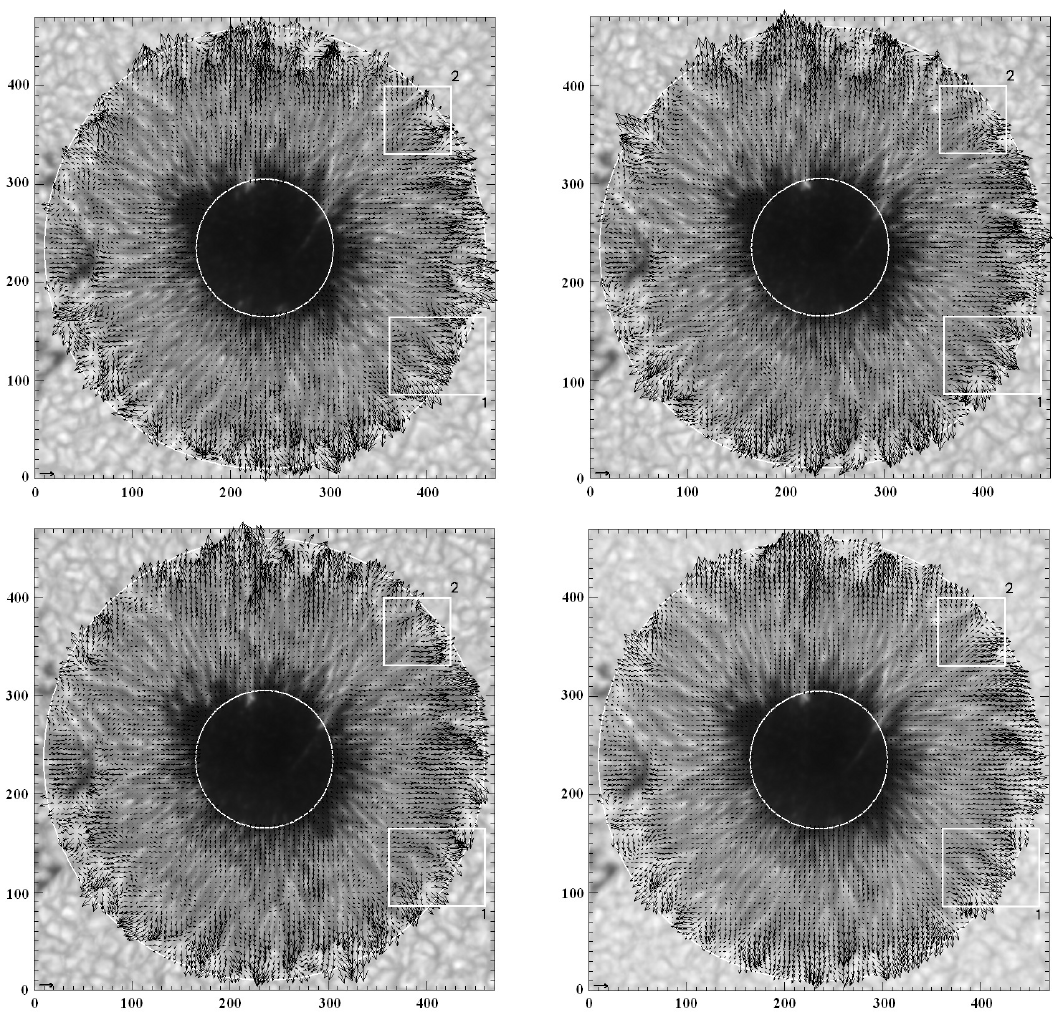}}
 \caption[]{Three successive vector flow fields and their mean.
 The whole time interval (48~min) is divided in three consecutive sequences (16~min)
 to determine these three maps choosing a Gaussian-weighted correlation window with
 a FWHM of 1~arcsec: upper-left is for the first 16~min, upper-right is for the middle 16~min
 and lower-left is for the last 16~min. Lower-right map shows the average of these three successive vector flow fields.
 The backgrounds are the corresponding time-averaged intensity maps.
 White circles are the defined inner and outer borders of the penumbra.
 The horizontal arrows at the lower left corners are equivalent to 1~km/s.
 Two while boxes show two selected filamentary flows. See the text for a detailed
 study.
 }
 \label{3vecflow}
\end{figure}

The whole time interval (48~min) is now split in three windows of
16~min resulting in three consecutive sequences which are used to
determine three new vector flow fields choosing a Gaussian-weighted
correlation window with a FWHM of 1~arcsec. The results are shown in
Fig.~\ref{3vecflow}. Also, the average of these three vector flow
fields is illustrated in Fig.~\ref{3vecflow}, lower-right map, and
should be similar to the one in Fig.~\ref{gvecflow}. The precision
of the speeds is estimated to be 0.08~km/s. These three subsequent
maps show slow flows up to 0.85, 0.85 and 1.0~km/s, respectively,
with an average of 0.2~km/s in the inner penumbra. However, in the
outer penumbra, the flows reach the maxima of 2.4, 2.7 and 3.0~km/s
in the subsequent maps, respectively, with the same mean speed of
about 0.4~km/s. The temporal changes of speeds as well as directions
of proper motions are seen all over the penumbra, especially at the
periphery of the penumbra.

\begin{figure} [p!] 
 \vspace*{50pt}
 \centerline{\includegraphics[width=\textwidth,clip=]{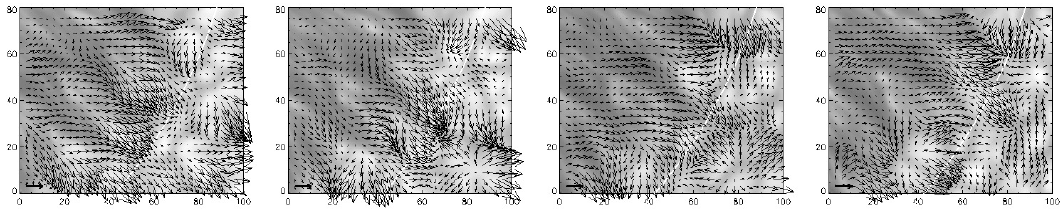}}
 \caption[]{Four successive vector flow fields similar to Fig.~\ref{3vecflow} but for the lower-right portion of the penumbra
 (box~1 in Fig.~\ref{3vecflow}) focusing on a converging filamentary flow (along the diagonal of the map) at the outer penumbra.
 The left map is the first in time. The backgrounds are the corresponding time-averaged intensity maps.
 The white curve is the defined boundary of the outer penumbra.
 }
 \label{box1vecflow}

 \vspace*{60pt}

 \centerline{\includegraphics[width=\textwidth,clip=]{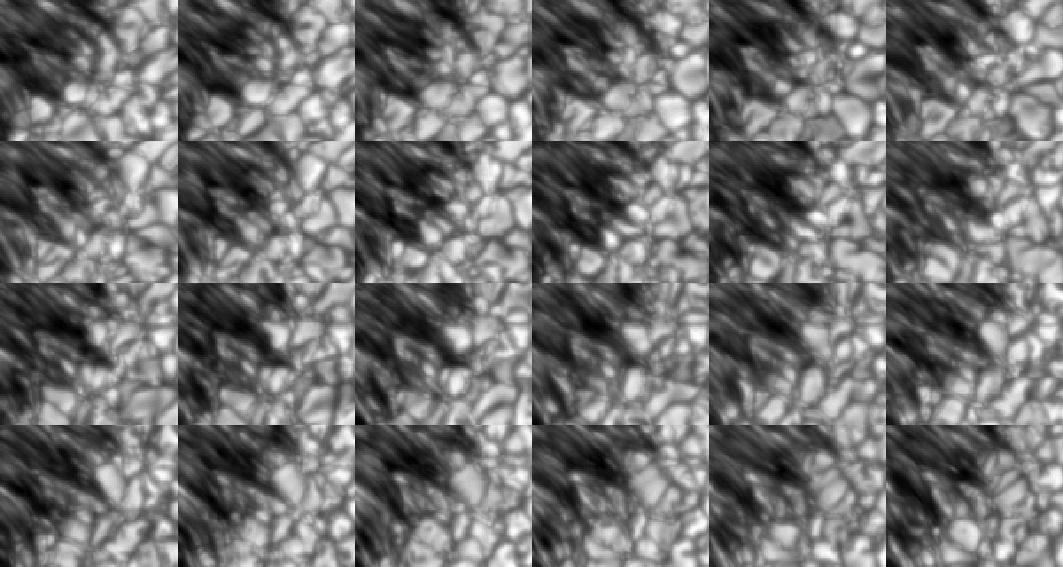}}
 \caption[]{The successive intensity images with a 2~min lag illustrates the evolution of brightness structure of
 the region confined by box~1 in Fig.~\ref{3vecflow}. The upper-left image is the first in time and the lower-right one is the last.
 Each row, from top to bottom, corresponds to the each map in Fig.~\ref{box1vecflow}, from left to right, respectively.
 }
 \label{box1int}
\end{figure}

\subsubsection{Evolving Filamentary Flows at outer penumbra}\label{evolfilament}

As mentioned before, despite of the circular shape of the sunspot,
the vector proper motions are not isotropic in three successive maps
displayed in Fig.~\ref{3vecflow}. At the outer penumbra, the proper
motions tend to diverge along filamentary structures with outward
motions, resulting in adjacent convergent filamentary structures as
well \cite{marq06,denker08}, resembling roll convection
\cite{zakharov08} while inner penumbra does not display this
behavior, probably because of the dominant radial migration of
penumbral grains, if there exists.

For a detailed study, four successive vector flow fields of the
lower-right portion of the outer penumbra (the box~1 in
Fig.~\ref{3vecflow}) are displayed in Fig.~\ref{box1vecflow}. These
maps are focusing on a converging filamentary flow (caused by two
adjacent diverging flows) at the outer penumbra which develops along
the diagonal of the map. Fig.~\ref{box1int} shows the corresponding
successive intensity images with a 2~min lag illustrating the
evolution of the brightness structure of the region. Each row, from
top to bottom, corresponds to each map in Fig.~\ref{box1vecflow},
from left to right, respectively.

Comparing Figs.~\ref{box1vecflow} \& \ref{box1int}, we can see that
the converging filamentary flow occurs in a dark radial channel and
the diverging flows are formed by the evolution of thin bight
fibrils. Also, the large speeds at the penumbra boundary are
produced by the displacement and/or the fragmentation of the bright
fibrils. It seems that these fast proper motions are sharply
interrupted (see the centre of the first map in
Fig.~\ref{box1vecflow}) by emerging a divergence center that is
related to a multi-sector granule (see intensity images in
Fig.~\ref{box1int}) forming a sink/convergence point in between
(lower-right corner of the second and third map in
Fig.~\ref{box1vecflow}). This behavior is repeated in the forth map
in Fig.~\ref{box1vecflow} when two new divergence centers evolve.
Also, it seems that the evolution and fragmentation of bright
fibrils are closely related to the formation and evolution of the
multi-sector granules connected to them. A multi-sector granule is
the expanding and splitting granule producing a strong divergence
center, so-called ``Rosetta" by Sobotka et al.  \cite{sob99b}, in
LCT maps which can be a member of a ``Tree of Fragmenting Granules"
or simply of a ``family of granules" reported by Roudier et al.
(\cite{roud03} and references therein) and Bonet et al.
\cite{bonet05}, respectively.

\begin{figure} [p!] 
 \vspace*{40pt}
 \centerline{\includegraphics[width=\textwidth,clip=]{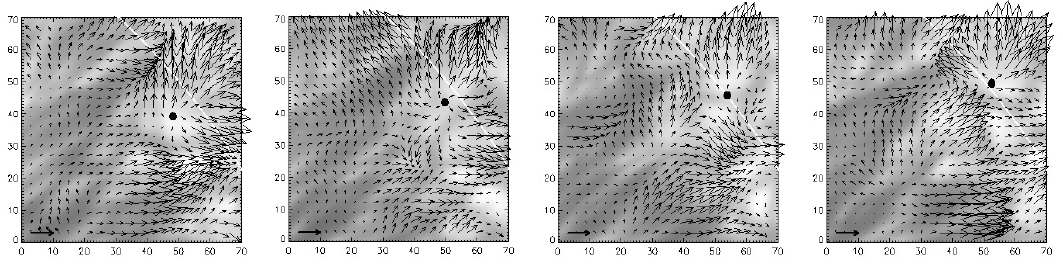}}
 \caption[]{Similar to Fig.~\ref{box1vecflow} but for the upper-right portion of the outer penumbra
 (box~2 in Fig.~\ref{3vecflow}) focusing on a granular flow (divergence center marked by black dot) at
 the periphery of the penumbra that is pushed away by the evolution of a diverging filamentary flow.
 }
 \label{box2vecflow}

 \vspace*{50pt}

 \centerline{\includegraphics[width=\textwidth,clip=]{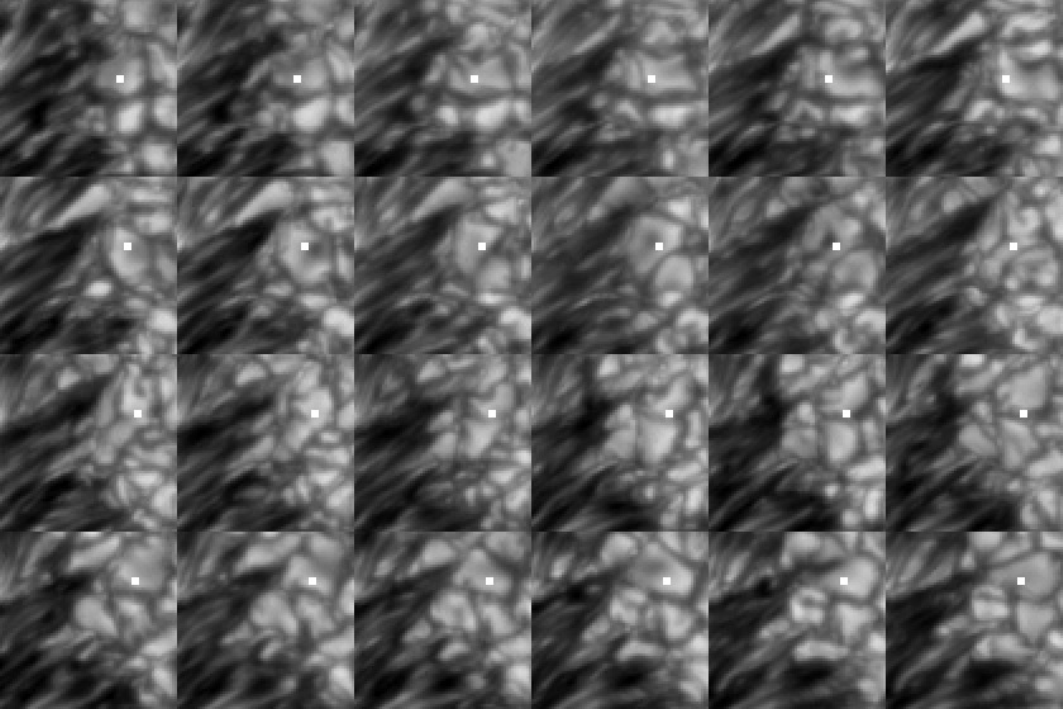}}
 \caption[]{Similar to Fig.~\ref{box1int} but for the region defined by box~2 in Fig.~\ref{3vecflow}.
 Each row, from top to bottom, corresponds to the each map in Fig.~\ref{box2vecflow}, from left to right, respectively.
 White dots mark the centers of the corresponding ``divergence center" shown in Fig.~\ref{box2vecflow}.
 }
 \label{box2int}
\end{figure}

Similar to Fig.~\ref{box1vecflow} but for the upper-right portion of
the penumbra (box~2 in Fig.~\ref{3vecflow}), four successive vector
flow maps focusing on a divergence center at the periphery of the
penumbra are shown in Fig.~\ref{box2vecflow} in which the divergence
center is marked by a black dot. Fig.~\ref{box2int} shows the
corresponding successive intensity images in which the white dots
mark the divergence center. A multi-sector granule is emerged and
the connected bright fibrils grow and produce a strong radial flow.
It is evident that the divergence center is not co-spatial with the
center of the developing multi-sector granule in the intensity
images. These maps illustrate how a divergence center is pushed away
with a mean speed of 0.6~km/s by the development of a diverging
filamentary flow produced by the evolution of a few long bright
fibrils (see Fig.~\ref{box2int}, especially the third row). Other
examples of this behavior can be seen around the penumbra.

\subsubsection{Zero-Flow Ring}\label{zfr}

\begin{figure} [p!]
 \centerline{\includegraphics[width=\textwidth,clip=]{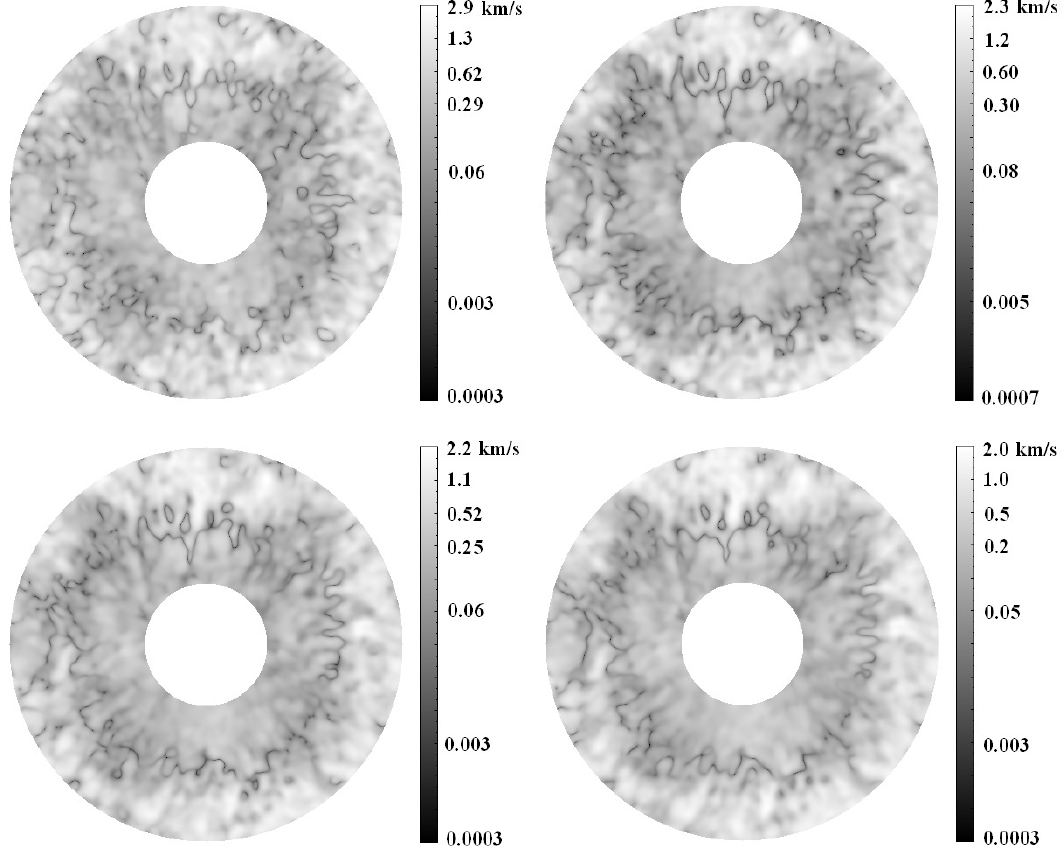}}
 \caption[]{Maps of the speeds of proper motions in logarithmic scale by choosing a Gaussian correlation
 window with a FWHM of 1~arcsec. Upper-left is for the first 12~min, upper-right is for the first 24~min and
 lower-left is for the first 36~min. lower-right map shows the map of the speeds of whole time series,
 the same as shown in Fig.~\ref{histspeed}.
 }
 \label{zfr_changes}
\end{figure}

To investigate the changes of ZFR as well as the divergence line in
time, we compute the maps of the speed of proper motions using the
first 12~min, then the first 24~min, and finally the first 36~min
and compare them with each other, and with the map of the speed of
the whole (48~min) time series shown in Fig.~\ref{gspeed}. These
four maps are illustrated in Fig.~\ref{zfr_changes}. These four maps
show slow flows with averaged speeds of 0.24, 0.20, 0.19 and
0.19~km/s in the inner penumbra, respectively. However, in the outer
penumbra, the averaged speeds of the flows reach 0.45, 0.40, 0.38
and 0.37~km/s in the maps, respectively. The smoothed speed obtained
in a flow map depends on the length of the time series over which
the LCT is applied. This is the reason that LCT algorithm gives
smaller speeds when applied over the whole time series. However, the
averaged speeds reach a convergent value during 48~min averaging.

It is evident in Fig.\ref{zfr_changes} that while a longer time
series of images is used (many subsequent images are included) to
compute the speed of proper motions, the ZFR establishes its shape:
the ZFR of this sunspot penumbra, with a low activity level, shows
stable backward/forward displacements \cite{denker08} or
discontinuities along itself during the 48~min observation.

We make a visual inspection on the evolution of penumbral filaments
to find out a possible relation between significant elongating
displacements of ZFR or its discontinuities and the intruding
filaments into umbra (compare the backgrounds in Fig.~\ref{3vecflow}
with the shape of ZFR in Fig.~\ref{gspeed}). However, in agreement
with Denker et al. \cite{denker08}, in one case the ZFR co-spatial
with the divergence line moves (is carried) toward umbra where
coordinated penumbral filaments intrude deeply into umbra, arrow~1
in Figs.~\ref{gvecflow} \& \ref{gspeed}. They have discussed that
this is the imprint of the cluster of extremely bright penumbral
grains (with a contrast in excess of 1.8 times the quiet Sun
intensity) on the penumbral flow field so that their speeds were
significantly increased in excess of 2~km/s. The applied time window
and correlation window in the LCT map of Denker et al.
\cite{denker08} were, respectively, 30~min and
$2.5\times2.5$~arcsec$^{2}$ contained a Gaussian weighting function
with a FWHM of 0.5~arcsec.

On the other hand, during some deeply intrusion, the ZFR is
``ruptured" or distorted (see Fig.~\ref{gspeed}, arrows~2~\&~4),
while the divergence line moves outward in this case (see
Fig.~\ref{gvecflow}; arrows~2~\&~4). Also, in some cases, e.g.
indicated by arrow~3 in Figs.~\ref{gvecflow} \& \ref{gspeed}, while
penumbral filaments are intruding into umbra, neither the divergence
line moves inward nor a discontinuity forms along the ZFR.

\section{Discussion and Conclusion}\label{conclusion}

Based on horizontal proper motions obtained from LCT, we confirm the
existence of a divergence line at the middle of penumbra
($r/R_{spot}\approx0.65$) dividing inward motions in inner penumbra
from outward motions in outer penumbra (see e.g.,
\cite{denker08,sob99a}). The displacement of the divergence line
towards umbra was reported by Denker et al. \cite{denker08}. They
suggested that a tight cluster of extremely bright penumbral grains
is the cause of this displacement where the speed of flow field
significantly increases. Our studied penumbra shows an exception
indicated by arrow~3 in Figs.~\ref{gvecflow} \& \ref{gspeed}.

Whereas the divergence line is continuous, a distorted ringlike
feature with very slow speeds ($\approx50$~m/s; ZFR) co-spatial with
the divergence line, is seen in penumbra. A few regions of the
penumbra show deeply intrusion of coordinated penumbral filaments.
The ZFR at some regions shows discontinuities showing relatively
considerable speeds of about 150~m/s there. In these regions, the
divergence line is not displaced toward umbra (arrows~2~\&~4 in
Fig.~\ref{gvecflow}). In accordance to Denker et al.
\cite{denker08}, in one case a considerable displacement of the ZFR
co-spatial with the divergence line towards the umbra is seen. Also,
stable local backward/forward shifts of ZFR are probably related to
the different evolutionary proper motions of adjacent penumbral
\textit{vena} (filaments). We must mention that the different
suitable values of FWHM do not significantly affect the shape of the
ZFR.

Bellot Rubio \cite{bell03} has defined the position at around
$r/R_{spot}\approx0.7$ penumbral radii as the transition between
inner and outer penumbra where the azimuthal averages of atmospheric
parameters, e.g. magnetic field inclination and filed strength
undergo significant changes. He has concluded that this behavior can
suggest the region at around 0.7 penumbral radii is the position
where a new family of penumbral flux tubes carrying the Evershed
flow emerges (see also \cite{plaza01,bell_etal03,bell06,trit04}).

On the other hand, Borrero \& Ichimoto \cite{bor_ich11} have
discussed about two distinct regions with two different magnetic
structures in a sunspot penumbra. A region corresponds to the inner
part of the penumbra, including the umbra $r/R_{spot}<0.5$, where
the total magnetic field strength decreases from the deep
photosphere upwards and the inclination of the magnetic field vector
remains constant with height. In the outer part, $r/R_{spot}>0.5$,
the situation, however, reverses. The total magnetic field strength,
as well as the vertical and horizontal components of the magnetic
field, increases from the deep photosphere to the higher
photosphere. In this region, the inclination of the magnetic field
vector decreases towards the higher photospheric layers (see also
\cite{plaza01}). Also, they have investigated the variation of the
azimuthally averaged plasma-$\beta$ (the ratio of gas pressure to
magnetic pressure) as a function of the normalized radial distance
in the sunspot. At continuum level, they have obtained
$\beta\simeq1$ in the inner part of the sunspot. However, $\beta$
increases from middle of penumbra outwards ($r/R_{spot}>0.6$) and
the magnetic field becomes highly non-potential where the dynamics
of the outer penumbra are strongly dominated by the plasma motions.
Therefore, it is not unlikely that the formation of ZFR (or
divergence line) is related to the existence of the two different
magnetic structures in a sunspot penumbra as discussed by, e.g.
Bellot Rubio \cite{bell03} or Borrero \& Ichimoto \cite{bor_ich11}.

On the other hand, Liu Yang et al. \cite{liu13} have carried out a
comparison between photospheric flows and subphotospheric flows
inside active regions. The photospheric flow field was obtained from
\textit{differential affine velocity estimator for vector
magnetograms} \cite{schuck08} using the data of HMI observed vector
magnetograms. The subphotospheric flow field in a shallow depth down
to 1 Mm was inferred from time-distance analysis using the HMI
observed Doppergrams. In their work, a mature and simple active
region (NOAA~11084) without significant flaring activities shows
inward and outward flows in both photospheric and subphotospheric
layers of the sunspot. Separation of the inward- and outward-flows
(divergence line) takes place in the sunspot penumbra. However, the
divergence lines are different in these two layers: the area having
inward subphotospheric flow was larger than that in the photosphere.
Also, they concluded that the horizontal flows caused by flux
emergence do not extend deeply into the subsurface.

Also, similar to the findings of Westendorp Plaza et al.
\cite{plaza01}, in the radial intensity profile of the studied
penumbra displayed in Fig.~\ref{IV_vs_r}, we find a weak bright ring
close to the ZFR ($r/R_{spot}\approx0.65$) and a wide plateau region
from 0.6 to $0.8R_{spot}$ . On the other hand, Rempel
\cite{rempel12} has presented a series of numerical sunspot models
in which they studied the dependence of the resulting sunspots on
the magnetic top boundary condition and grid resolution by keeping
initial condition fixed. In their work, only the simulated sunspots
with higher resolutions show the formation of a plateau like feature
at normalized intensity around 0.7 in the intensity map of the inner
penumbra at around $0.6-0.8R_{spot}$. However, the average of umbral
intensity is about half of the umbral intensity of their simulated
sunspots.

The diverging filamentary flows with outward motions in outer
penumbra (Fig.~\ref{3vecflow}) are consistent to the behavior of the
penumbral filaments simulated by Rempel \cite{rempel12}. The top
view of the penumbral filaments in his numerical simulations shows a
central upflows with adjacent lateral downflows (see also
\cite{zakharov08}). While plasma is moving outward (Evershed flow),
it moves to the edges of the filaments where it turns back into the
photosphere, i.e. individual flow elements consist of both a radial
outflow and lateral overturning motions.

The proper motions around the sunspot obtained by selecting high
resolution integration parameters in LCT (1~arcsec and 12~min;
Figs.~\ref{box1vecflow} \& \ref{box1int}) can be understood by
combination in a vector summation of two components, as described in
\cite{bonet05}: both mesogranular/convective flows producing
divergence centers, and a large-scale flow driving the divergence
centers.

However, as described in Sect.~\ref{evolfilament}, at the near
vicinity of the penumbra, in surrounding granulation some divergence
centers are strongly pushed away as a whole with an average speed of
about 0.6 km/s by a developing filamentary flow so that, after a
while, we observe only a locally outward flow on those
already-diverging centers (compare the second map in
Fig.~\ref{box2vecflow} with the third and fourth maps). In other
words, in a vector summation the filamentary flow cancels out the
component of the diverging flow on the penumbral side of a
divergence center and increases its opposite component. Similar
phenomenon has been reported in \cite{deng07} where the moat flow
component affects convective granular velocities around a pore
located far from solar disc center ($\mu=0.43$). However, Bonet et
al. \cite{bonet05} have reported only elongated divergence centers
(mesogranules as called by them) whose diverging speeds away from
the sunspot are higher than those directed towards the sunspot. At
present, we could not conclude if there is any relation between the
mentioned large-scale flow and the evolving filamentary flows in
outer penumbra.

At the outer penumbra, the converging filamentary flow occurs in a
dark radial channel and the diverging flows are formed by the
evolution of thin bight fibrils. Also, the large speeds at the
penumbra boundary are produced by the displacement and/or the
fragmentation of the bright fibrils. Furthermore, it seems that the
evolution and fragmentation of bright fibrils are closely related to
the formation and evolution of the multi-sector granules connected
to them in their near surrounding granulation.

It is needed to investigate the temporal variations of the
divergence line as well as the ZFR using a long time series of
images to find out the behavior of the the two proposed magnetic
structures \cite{bor_ich11,bell03} and their competition in
formation and evolution of a sunspot penumbra.

\section*{Acknowledgment}
\textit{Hinode} is a Japanese mission developed and launched by
ISAS/JAXA, collaborating with NAOJ as a domestic partner, NASA and
STFC (UK) as international partners. Scientific operation of the
\textit{Hinode} mission is conducted by the \textit{Hinode} science
team organized at ISAS/JAXA. This team mainly consists of scientists
from institutes in the partner countries. Support for the
post-launch operation is provided by JAXA and NAOJ (Japan), STFC
(UK), NASA, ESA, and NSC (Norway). The author would also like to
thank M. Sobotka and the referee for suggestions that improved the
paper.


\end{document}